\documentclass{elsart}
\usepackage{graphicx}
\usepackage{amssymb}

\begin{document}

\begin{frontmatter}

\title{Measurement of Pressure Dependent Fluorescence Yield of Air: 
Calibration Factor for UHECR Detectors} 

\author[umt]{J.W.~Belz,}
\author[utah]{G.W.~Burt,}
\author[utah]{Z.~Cao,}
\author[cospa]{F.Y.~Chang}
\author[cospa]{C.C.~Chen}
\author[cospa]{C.W.~Chen}
\author[slac]{P.~Chen,}
\author[slac]{C.~Field,}
\author[utah]{J.~Findlay,}
\author[utah]{P.~H\"{u}ntemeyer\corauthref{cor1},}
\author[cospa]{M.A.~Huang}
\author[cospa]{W-Y.P.~Hwang}
\author[slac]{R.~Iverson,}
\author[utah]{B.F Jones,}
\author[utah]{C.C.H.~Jui,}
\author[umt]{M.~Kirn}
\author[cospa]{G.-L. Lin}
\author[utah]{E.C.~Loh,}
\author[utah]{M.M.~Maestas,}
\author[utah]{N.~Manago,}
\author[utah]{K.~Martens,}
\author[utah]{J.N.~Matthews,}
\author[slac]{J.S.T.~Ng,}
\author[slac]{A.~Odian,}
\author[slac]{K.~Reil,}
\author[utah]{J.D.~Smith,}
\author[utah]{R.~Snow,}
\author[utah]{P.~Sokolsky,}
\author[utah]{R.W.~Springer,}
\author[utah]{J.R.~Thomas,}
\author[utah]{S.B.~Thomas,}
\author[rutgers]{G.B.~Thomson,}
\author[slac]{D.~Walz,}
\author[rutgers]{and A.~Zech}
\collaboration{The FLASH Collaboration}
\address[umt]{University of Montana, Department of Physics and Astronomy,
Missoula, MT~59812, USA.}
\address[utah]{University of Utah,
Department of Physics and High Energy Astrophysics Institute,
Salt Lake City, UT~84112, USA}
\address[cospa]{Center for Cosmology and Particle Astrophysics, Department
of Physics, National Taiwan University, 1 Roosevelt Road , Section 4, 
Taipei 106-17, Taiwan}
\address[slac]{Stanford Linear Accelerator Center,
2575 Sand Hill Road, Menlo Park, CA~94025, USA} 
\address[rutgers]{Rutgers --- The State University of New Jersey,
Department of Physics and Astronomy,
Piscataway, NJ~08854, USA}
\corauth[cor1]{
Corresponding~author.\ {\it E-mail~address}:~petra@cosmic.utah.edu
}

\newpage

\begin{abstract}

In a test experiment at the Final Focus Test Beam of the Stanford Linear 
Accelerator Center, the fluorescence yield of 28.5 GeV electrons in air 
and nitrogen was measured. The measured photon yields between
300 and 400 nm at 1~atm and 29$^{\circ}$C
are
\begin{eqnarray}
  Y(760~{\rm Torr})^{\rm air}= 4.42 \pm 0.73 
  {~~~\rm and~~~}
  Y(760~{\rm Torr})^{\rm N_2}= 29.2 \pm 4.8 
  \nonumber
\end{eqnarray}
photons per electron per meter.
Assuming that the fluorescence yield is proportional to the energy deposition
of a charged particle traveling through air, good agreement with measurements 
at lower particle energies is observed.

\end{abstract}

\begin{keyword}

Nitrogen fluorescence \sep Air fluorescence \sep Extensive air shower 
\sep Ultrahigh-energy cosmic rays 

\PACS 95.55.Vj \sep 96.40.Pq \sep 96.40.De \sep 32.50.+d 

\end{keyword}
\end{frontmatter}

\section{Introduction}
\label{sect-intro}
Measuring the energy spectrum of Ultra High Energy Cosmic Rays (UHECR) is 
the goal of several past, present, and future experiments using the 
air fluorescence technique. 
It is of special interest to measure the spectrum in the range of $10^{20}$~eV 
and establish whether it is suppressed as predicted by the GZK mechanism \cite{GZK}.
The published results of the two, at that time largest, 
experiments collecting UHECR data disagree \cite{agasa-hires}. One of those
detectors, the High Resolution Fly's Eye (HiRes) utilizes air fluorescence
to detect cosmic rays and to determine their energy. The other experiment, the
Akeno Giant Air Shower Array (AGASA),
a ground array of scintillation counters in Japan, used the particle sampling
technique. The energy dependent HiRes flux measurement is systematically 
smaller than that of AGASA and is consistent with a GZK suppression. 
One systematic uncertainty in the case of the fluorescence technique is the 
uncertainty of the measured fluorescence yield of charged particles in air 
itself. A more precise study of the yield 
would be valuable in seeking the cause for
the apparent discrepancies between the techniques.

There are a number of previous fluorescence yield
experiments. In his thesis from 1967 Bunner summarized the existing data and 
quoted uncertainties of around 30\% on the reported fluorescence efficiencies 
\cite{Bunner:1967}.
Bunner's work served as the standard reference for fluorescence technique 
based cosmic ray experiments into the nineties.
In a more recent experiment \cite{Kakimoto:1996}, 
Kakimoto {\it et al.}~measured
the total fluorescence yield between 300 and 400 nm with an uncertainty 
of $>$10\% and the yield at three pronounced lines - 337~nm, 357~nm, and 
391~nm. The newest measurements using a $^{90}\mathrm{Sr}~\beta$ source 
published by Nagano {\it et al.}~\cite{Nagano:2003,Nagano:2004}, determined 
the total fluorescence yield between 300~nm and 406~nm with a systematic 
uncertainty of 13.2\% as the sum of the yields of 15 wave bands which were 
measured separately using narrow band filters.
An improvement in the present level of accuracy and confidence is necessary, 
from measurements not subject to the same set of systematic uncertainties, especially since other systematics,
like the atmospheric uncertainty, which depends on 
variable pressure profile, transparency, scattering, etc., 
are expected to be reduced significantly in the near future.

New experiments using the fluorescence technique are
beginning to take data, are in construction, or are being planned. These 
include 
the hybrid detectors of the Pierre Auger Observatory~\cite{auger}, and
of the Telescope Array (TA)~\cite{ta:2004},
or the space-based fluorescence detectors 
EUSO~\cite{euso} and OWL~\cite{owl}.
These experiments are designed to increase the detection aperture 
and statistics in the ultra high energy region.
The hybrid detectors
should also help to resolve the disagreements between the particle sampling
technique and the fluorescence technique. Independent measurements of the 
fluorescence yield of charged particles in air as presented in this paper, and refinements proposed to follow this work, will complete the picture. 

The test experiment presented in this paper, T-461, was
conducted at the Stanford Linear Accelerator 
Center (SLAC) to study the feasibility of a larger fluorescence experiment, 
FLASH\footnote{FLuorescence in Air from SHowers}. Both experiments have since
been installed in SLAC's Final Focus Test Beam (FFTB) tunnel. 
FLASH aims to measure the net fluorescence yield as well as the yields of the 
individual spectral lines with a systematic uncertainty of less than 10\%.
Measurements with mono-energetic electrons, and separately with 
electron-positron 
showers downstream of thick materials, are used to measure the  
fluorescence yield down to an electron energy of 100 keV\cite{e-165}. 
T-461 used only the mono-energetic beam approach 
and was designed to measure the total 
fluorescence yield between 300 and 400 nm using a UV bandpass
filter as installed in 
the HiRes experiment. In the following Section, the experimental setup of 
T-461 will be described in detail. In Section 
\ref{sect-selection}, the selection of good quality data is described. This 
is followed by a brief description of the calibration of the experimental 
setup in Section \ref{sect-un-cal}. The fluorescence yield measured in air 
and nitrogen is presented in Section \ref{sect-results} along with a list 
of the systematic errors which were studied. In the last Section, the T-461 
results are compared with previous measurements and improvements are discussed
for the full scale fluorescence experiment FLASH.

\section{Experimental setup}
\label{sect-detector}
The experiment T-461 was carried out in the Final Focus Fest Beam
at SLAC. It was installed in an
air gap of the beam pipe, approximately 35 cm long, downstream of the dump
magnets. The beam is focused effectively at infinity in this region, so that
the bremsstrahlung from the beam windows and thin experimental equipment in
this region continues along the electron beam to the dump. In addition, since
the thickness of the material in the beam was held below 1\% of a 
radiation length, multiple scattering of beam particles into downstream
collimators is negligible. 

A beam spot size of about 2$\times$1~mm was generated in this region,
with intensities between $10^9$ and 2 $\times 10^{10}$ e$^-$/pulse. 
Between $10^8$ and $10^9$ electrons per pulse, the 
intensity is below the sensitivity of the beam position monitors and 
feedback becomes inoperative. Nonetheless, intensities as low as $10^8$ were 
delivered for 24 hours during T-461. However, the beam current measuring 
toroidal ferrite-core current transformer
registered intensity variations of $\approx$ 30\% during this period. 
The toroid was calibrated using charge-injection on a one-turn winding. The
accuracy was determined to be 10\% for beam charges below 10$^9$ e$^-$ by 
cross comparison with a high-accuracy toroid during a subsequent FLASH 
run.\footnote{Details of the calibration of the high-accuracy toroid is the 
subject of a future publication.}

\begin{figure}
  \includegraphics[angle=-90.,width=\columnwidth]{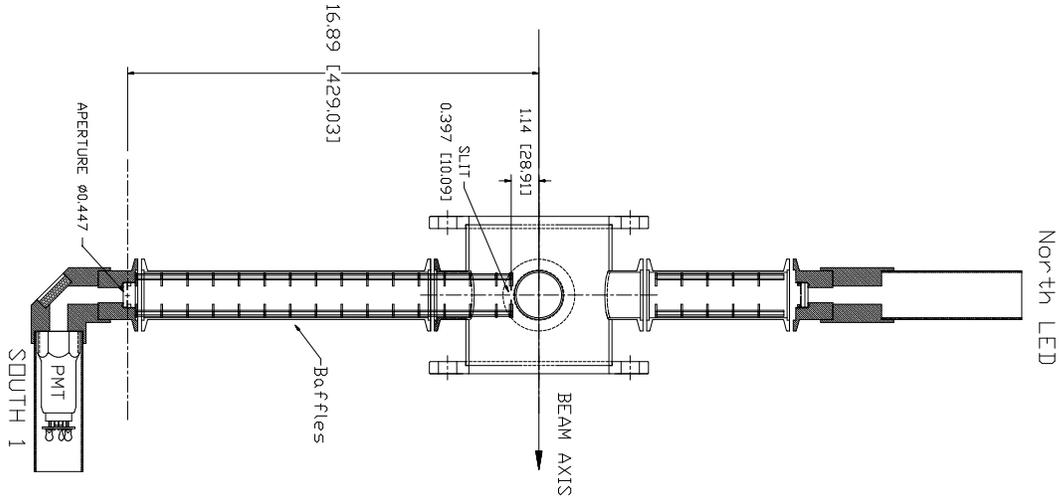}
  \caption{The experimental setup showing one of two orthogonal
optical pathways through baffled channels, and 45 degree reflection
to a photomultiplier tube (PMT). The values are given in inches [mm].}
  \label{fig-illustration}
\end{figure}

The thin target installed in FFTB during T-461 was a 1.6 liter 
air-filled cylinder, coaxial with the beam. The vessel had 
thin beam windows and 
radial ports to allow light to reach shielded photomultiplier tubes (PMTs) 
as shown in Figure~\ref{fig-illustration}.  The inside of the vessel was 
black-anodized and baffled 
to suppress all but direct light from the beam. 
The fluorescence light was detected in two independent radial tubes. In each, 
the optical aperture was defined by a slit near the beam axis, measured to be 
1 cm parallel, and 1.7 cm perpendicular to it. In each tube, light passed 
through a series of circular apertures forming a baffle.
At the far end of the 42.9~cm long baffle,
the PMTs 
were protected from scattered 
background radiation by using dielectric mirrors to reflect the  
fluorescence light through 90$^0$ into the detector.
There, the PMTs were encased in a lead vault.
As can be seen in Figure~\ref{fig-illustration}, a UV band pass filter 
from the High Resolution Fly's Eye (HiRes) experiment was 
installed between the dielectric mirror and PMT.
The PMTs, Philips (now Photonis) XP3062/FL, are also the type used in the 
HiRes experiment.
The optical unit including the mirror, UV filter, and PMT
was calibrated at the University of Utah after 
T-461 data taking was completed. 

To monitor the stability of the PMT response in-situ, ultraviolet LEDs were 
installed on arms opposite the detector arms. The LEDs were fired 
between beam pulses emitting light across the fluorescence cylinder. 
To measure beam induced backgrounds in the PMTs, optically hooded 
--- ``blind'' --- PMTs were mounted beside the ``live'' tubes.  Signals of the 
blind tubes
were collected continuously along with the signals of the live tubes and
normalized to the live signals by collecting data while the fluorescence 
chamber was filled with a non-fluorescing gas.
 
The gas system allowed the flow of a premixed gas with a pressure in
the vessel between 3 and 760 Torr. The
pressure was set manually, but monitored by computer during data taking. 
Data for dry air and pure nitrogen were collected with flowing gas. 
Data for various air-nitrogen mixtures were collected without flowing
the gas through the chamber.

During T-461, a CAMAC based DAQ system was used to collect the data.
Pulse amplitudes from the PMTs and the beam toroid signal 
were recorded using LRS 2249W ADCs from Lecroy.
PMT high voltage, pressure and temperature were 
digitized with the Smart Analog Module, a 32-channel module used to 
digitize analog signals. 

\section{Data taking and event selection}
\label{sect-selection}
Altogether about 1 million events were recorded with a mean high voltage of 
1186 V supplied to the PMTs. The voltage was constrained to better than
half a volt over the entire run period.

Throughout the data taking, 
a special trigger, about once every 52 beam pulses, was used to measure 
ADC pedestals. 
The pedestal values were found to be stable around 45 and 47 counts for
the north and south signal PMTs, respectively.
From each beam event the nearest pedestal measurement was subtracted.

\begin{figure}
  \includegraphics[width=\columnwidth]{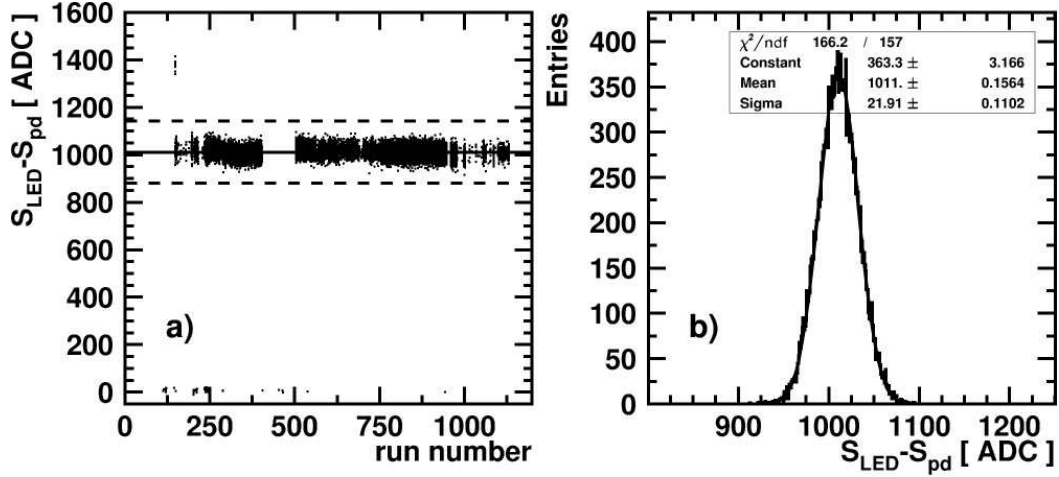}
  \caption{Pedestal subtracted response of the south signal PMT to the LED.  a) The south PMT response versus the run number.  The dashed lines represent the data quality preselection cut.  b) The projection of the response onto the signal axis.}
  \label{fig-LED_cut}
\end{figure}

The PMT response was tracked throughout the data taking with ultraviolet 
LEDs.  As for the pedestal events, LED events were also 
taken once every 52 beam 
events.  The response of the south PMT to the LED was very stable (around
2\%) during 
the experiment as seen in Figure~\ref{fig-LED_cut}a.  Similar to the 
pedestal subtraction scheme, each beam induced event was assigned the 
closest LED reading.

In order to remove poor quality data and data with unstable detector response, 
the LED data of the south PMT were fitted to a Gaussian, see 
Figure~\ref{fig-LED_cut}b.  Beam data during time periods where LED events 
were outside of 
$\pm6\sigma$ of the fitted mean were removed from the data set. The 
$\pm6\sigma$ band is represented by two dashed lines in 
Figure~\ref{fig-LED_cut}a. Less than 1.5\% of the data were excluded by
this requirement (see Table~\ref{table-cuts}).
\begin{table}
\begin{center}
\begin{tabular}{|lr|} \hline
\multicolumn{1}{|c}{Cut} &
\multicolumn{1}{c|}{Efficiency (\%)} \\ \hline
LED requirement & 98.6 \\
PMT correlation & 98.5 \\
Beam distribution & 84.7 \\
Linearity cut & 41.9 \\ \hline
Air &  15.8\\
Mixtures & 13.7 \\
N$_2$ (flowing) & 2.6  \\ \hline
\end{tabular}
\end{center}
\caption{Efficiency of the main preselection requirements imposed on the T-461 data.  See Section~\ref{sect-selection} for details.}
\label{table-cuts}
\end{table}

\begin{figure}
  \begin{center}
  \includegraphics*[width=8cm]{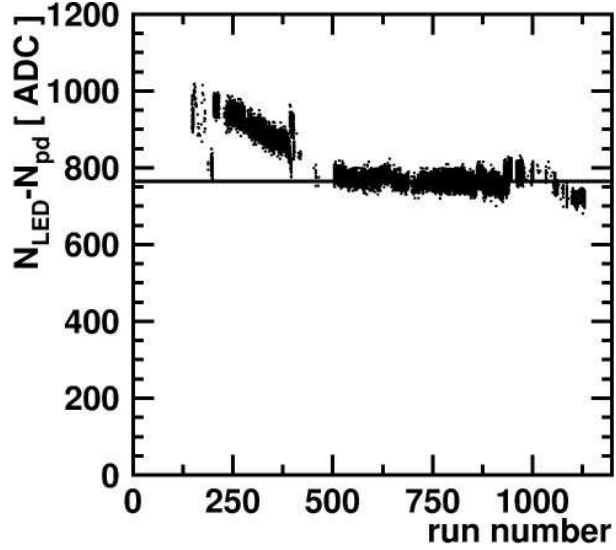}
  \end{center}
  \caption{Pedestal subtracted response of the north signal PMT to the LED.  The $x$-axis is the run number and the $y$-axis is the pedestal subtracted LED signal. }
  \label{fig-LED_dist}
\end{figure}

Figure~\ref{fig-LED_dist} shows that the north PMT's response was unstable,
probably due to a HV connector problem. 
While the north PMT data were used to check the data quality,
because of such wide systematic wandering, the north PMT 
was not used in the 
final result.  The pedestal subtracted LED 
signals were used to correct for gain changes shown in 
Figures~\ref{fig-LED_cut} and \ref{fig-LED_dist} by the formulas,
\begin{eqnarray}
S_{cd}  =  \left(S - S_{pd}\right)\times\frac{\langle S_{LED}\rangle}{S_{LED} - S_{pd}}
\end{eqnarray}
and
\begin{eqnarray}
N_{cd}  =  \left(N - N_{pd}\right)\times\frac{\langle N_{LED}\rangle}{N_{LED} - N_{pd}}. 
\label{eq-corr}
\end{eqnarray}
Here,
$S_{cd}$ and $N_{cd}$ are the stability corrected and pedestal subtracted 
south 
and north PMT signals, $S$ and $N$ are the raw ADC readouts from the PMTs,  
$S_{pd}$ and $N_{pd}$ are the signals of the assigned pedestal events, 
$\langle S_{LED}\rangle$ and $\langle N_{LED}\rangle$ are the fitted 
means of the pedestal 
subtracted LED distributions shown for the signal PMTs by horizontal lines 
in Figure~\ref{fig-LED_cut}a and \ref{fig-LED_dist}, and 
$S_{LED}$ and $N_{LED}$ are the signals of the closest LED events.  
The mean for all the LED events collected by the south signal PMT is
$\langle S_{LED}\rangle = 1011$ counts.
The mean of $\langle N_{LED}\rangle = 765$ counts was calculated only 
from the longest period of north PMT gain stability which occurred
between runs 504 and 947.

\begin{figure}
  \includegraphics[width=\columnwidth]{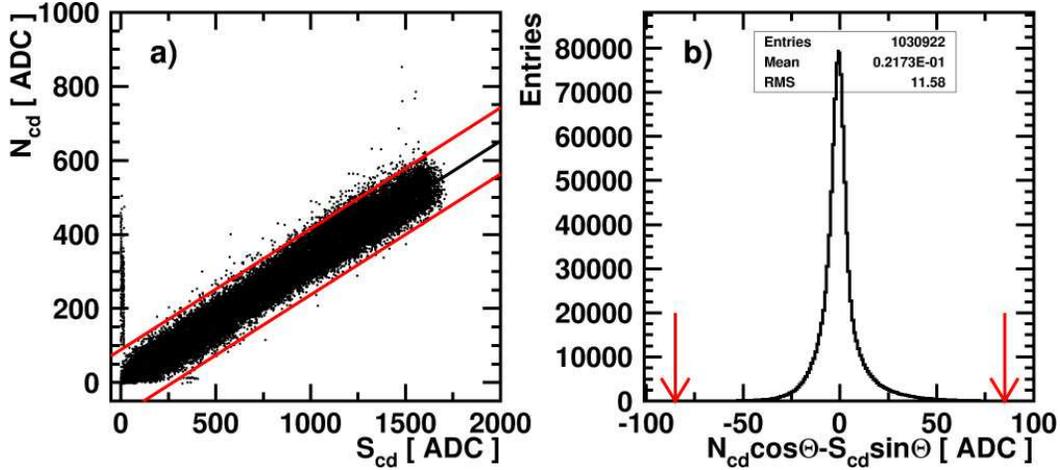}
  \caption{Signal PMT correlation.  a) The $x$-axis is the background subtracted and gain corrected south PMT signal and the $y$-axis is the background subtracted and gain corrected north PMT signal.  Air, N$_2$, and air-N$_2$ mixture events are shown.  The dashed lines correspond to the PMT correlation cut, see Table~\ref{table-cuts}.  b) Projection of the rotated distribution showing the correlation between $N_{cd}$ and $S_{cd}$ where $x = N_{cd}\cos\theta - S_{cd}\sin\theta$.  See text for details.}
  \label{fig-PMT-cut}
\end{figure}

After the PMT signals had been corrected for LED response the signal size was 
required to correlate well between the two signal PMTs.  In 
Figure~\ref{fig-PMT-cut}a the corrected signal, $N_{cd}$, of the north PMT 
is plotted versus the corrected signal, $S_{cd}$, of the south PMT.  A strong 
correlation for most of the data is quite apparent.  
The data were fitted to the linear function $N_{cd} = kS_{cd}$, 
$k$ being the correlation factor of the two signals. Then the axes were 
rotated so that the fitted line was 
the ordinate of a new plot.
The new 
abscissa is calculated according to the formula 
$x = N_{cd}\cos\theta - S_{cd}\sin\theta$, where $k = \tan\theta$.  
The rotated distribution was then projected onto the new abscissa.
The resulting distribution is shown in Figure~\ref{fig-PMT-cut}b.
All events 
beyond $\pm85.0$ ADC counts were removed from the data sample.  The 
upper and lower lines in Figure~\ref{fig-PMT-cut}a represent the 
170 ADC counts acceptance band.  The fraction of events remaining after 
this cut is listed in Table~\ref{table-cuts}.

\begin{figure}
\includegraphics[width=\columnwidth]{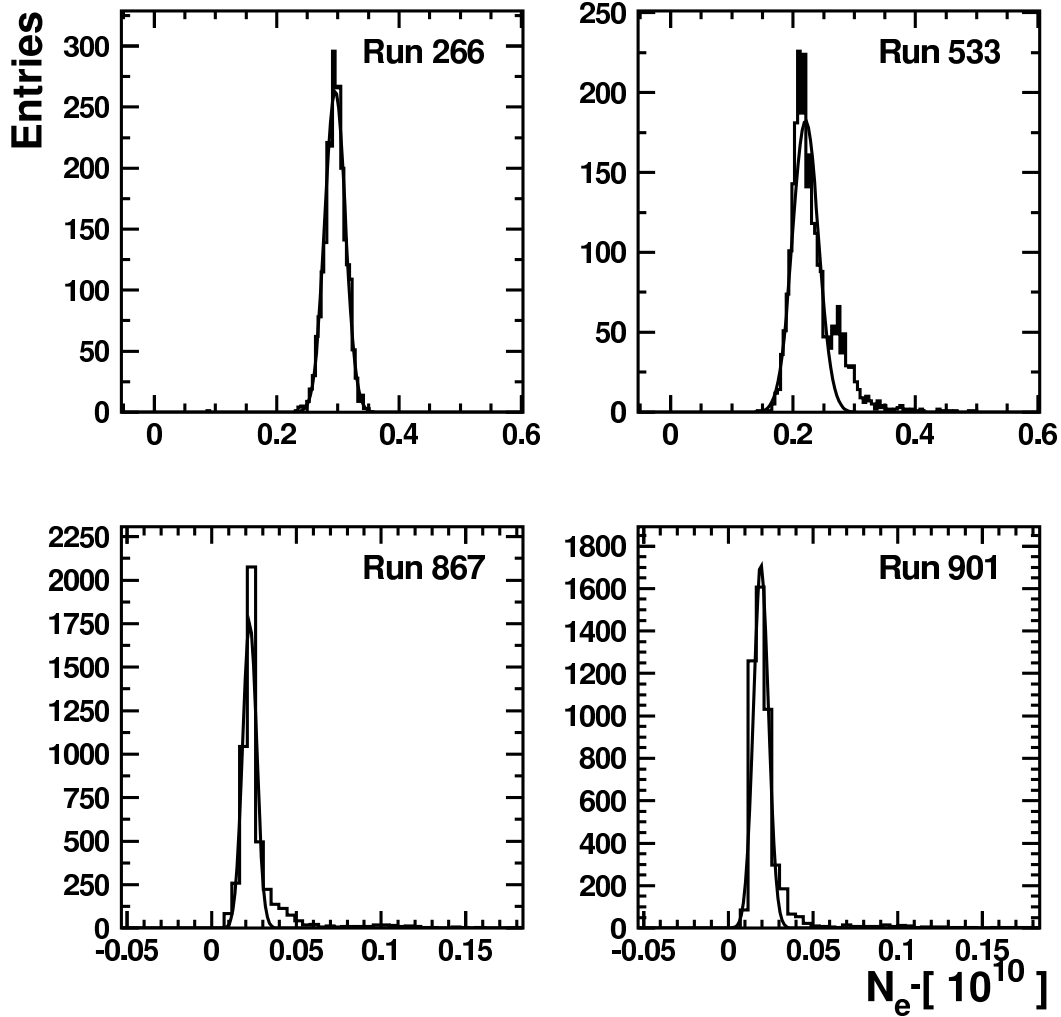}
  \caption{The beam charge distributions of four typical data runs as measured by
an upstream toroid during T-461.}
  \label{fig-beam-sample}
\end{figure}

During the two weeks of data collection, data were recorded at
beam charges between $10^{8}$ and $10^{10}\ e^-$ per bunch. 
For each data part, the target beam charge was specified, and
the charge of each 
bunch was measured with a toroid installed up-stream of the thin target 
vessel.  Figure~\ref{fig-beam-sample} shows example distributions of measured 
beam charges for four runs.  As can be seen, some of the distributions,
especially those with low beam intensity, have 
long tails possibly indicating poor beam quality. To remove the corresponding 
events from the data sample, the beam charge distribution of each run 
was fitted 
to a Gaussian curve and events lying outside of 2$\sigma$ were cut.  The 
selection efficiency of this requirement is also listed in
Table~\ref{table-cuts}.

\begin{figure}
  \includegraphics[width=\columnwidth]{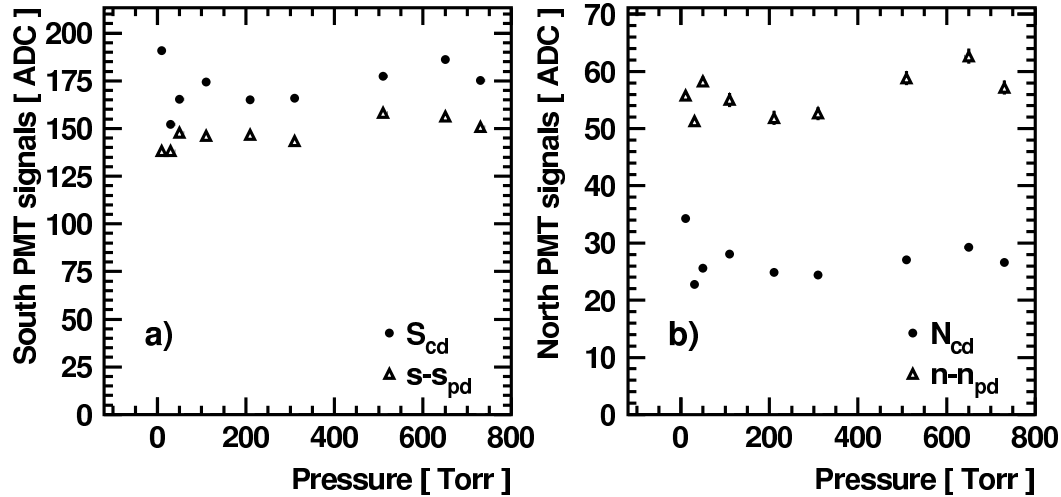}
  \caption{``Blind'' -- $s-s_{pd}$, $n-n_{pd}$ -- and ``signal'' -- $S_{cd}$, 
$N_{cd}$ -- PMT signals for ethylene at different pressures.  a) The data taken by the south PMTs and b) the north PMT data.  The relative difference between the ADC values illustrates the differences in gains between corresponding signal and blind PMTs.}
  \label{fig-bg-compare}
\end{figure}

In order to investigate possible background due to reflected Cherenkov light 
or other sources,
some data were collected while the gas chamber was filled with ethylene, 
a non-fluorescing gas. 
The same range of pressures were investigated as for air and nitrogen.  
The corrected signals, $N_{cd}$ 
and $S_{cd}$, of both PMTs for the ethylene data runs 
are plotted versus pressure in 
Figure~\ref{fig-bg-compare}.  The figure shows that $S_{cd}$ and $N_{cd}$ do 
not depend on the pressure, indicating negligible light background.  
The larger signal values for each signal tube at 
the lowest pressure probably indicate the presence of a trace amount of air.

\begin{figure}
  \includegraphics[width=\columnwidth]{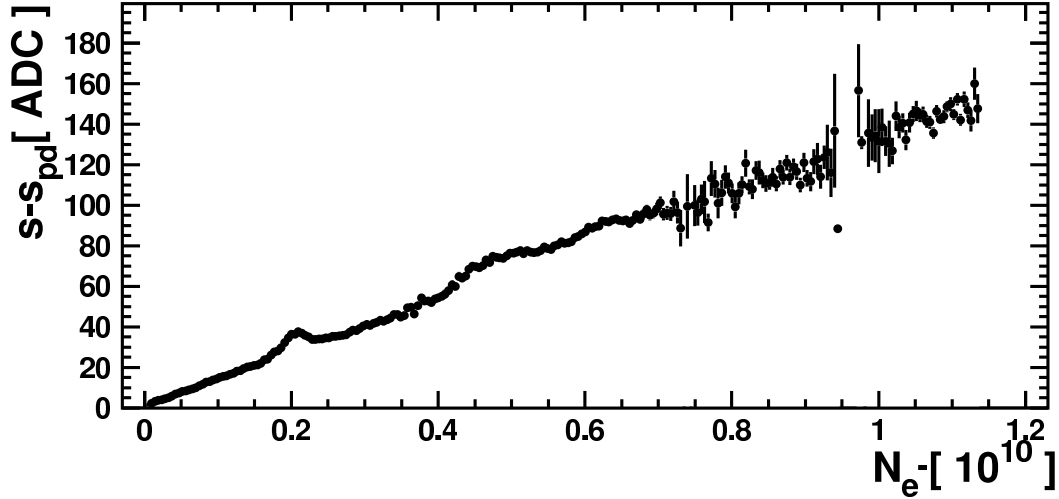}
  \caption{The south background PMT signal versus the number of beam electrons per bunch.}
  \label{fig-bg-beam}
\end{figure}

During the entire experiment, the beam related background was measured by two 
``blind'' PMTs positioned next to the ``signal'' PMTs.  
Figure~\ref{fig-bg-beam} shows that the blind PMT signal increases with the 
beam charge.  Using the ethylene data the blind PMT data was normalized to 
the signal PMT data.  The pedestal corrected signals of the blind PMTs are 
also plotted in Figure~\ref{fig-bg-compare}.  As can be seen, signal and 
blind PMTs track each other well.  The relative difference between the values 
illustrates the differences in gains between the signal and the blind PMTs.  
The mean ratios $\langle S_{bg}\rangle = 1.22$ and 
$\langle N_{bg}\rangle = 0.65$ are 
used for the beam related background subtraction:
\begin{eqnarray}
S_{fl} & = & S_{cd} - \langle S_{bg}\rangle (s - s_{pd})
\end{eqnarray}
and
\begin{eqnarray}
N_{fl} & = & N_{cd} - \langle N_{bg}\rangle (n - n_{pd}),
\label{eq-fl}
\end{eqnarray}
where $S_{fl}$, $N_{fl}$ are the background subtracted and LED corrected 
fluorescence signals in ADC counts, $s$, $n$ are the corresponding blind PMT 
signals, and $s_{pd}$, $n_{pd}$ are the assigned pedestal measurements for 
the blind photo tube channels.

\begin{figure}
\includegraphics[width=\columnwidth]{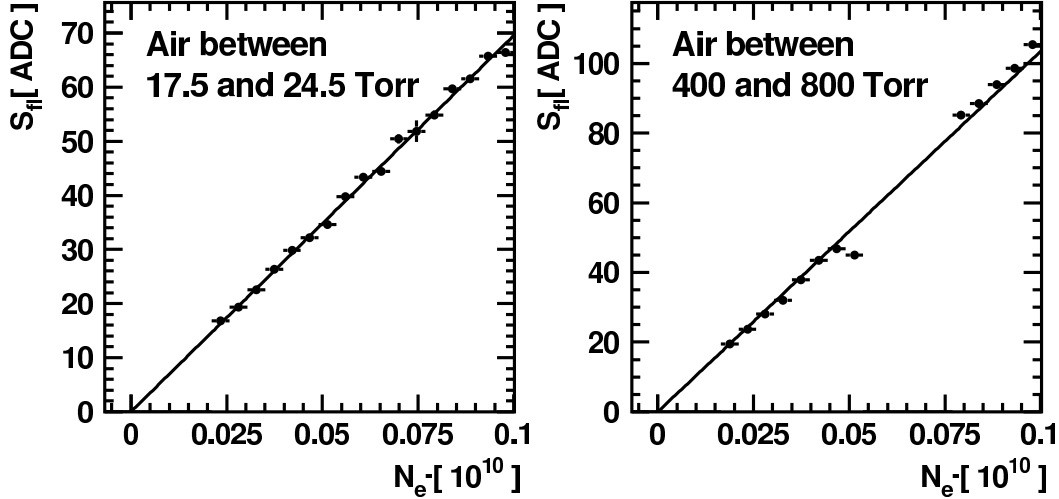}
  \caption{Stability corrected and background subtracted signal versus the number of electrons per beam pulse for two selected pressure regions for air.}
  \label{fig-signal vs beam nitrogen}
\end{figure}

During the run a non-linear enhancement of the fluorescence signal with 
respect to the beam charge was observed.  This effect was seen in both pure 
nitrogen and air.  This non-linearity is most likely caused by the 
acceleration of secondary electrons in 
the very strong radial electric field of the picosecond long beam pulse
of relativistic electrons.
This effect has, for example, been studied in 
\cite{Ng:2001is}.  For the purpose of measuring the fluorescence yield per 
electron, it is necessary to remove all data which were taken in the 
non-linear beam charge range.  In order to define an upper limit on the beam 
charge linear fits to the fluorescence signals $N_{fl}$ and $S_{fl}$ versus 
beam charge were performed for varying beam charge ranges. 
Based on the quality of 
the fits an upper limit of $1\times10^{9} e^-$ per beam pulse was chosen and 
all the data taken at higher beam charges were removed from the data sample.  
Two example fits with an applied beam charge limit of $1\times10^{9}~e^-$ per 
pulse are shown in Figure~\ref{fig-signal vs beam nitrogen}.

The final fraction of events remaining after all cuts for air, nitrogen and 
nitrogen-air mixtures are listed in Table~\ref{table-cuts}.  In the case of 
nitrogen, only data which were collected when the gas was flowing through the 
chamber were considered.  This ensured that potential impurities from small 
leaks of ambient air in the chamber were negligible.  The fluorescence yield 
$Y$ per electron per meter in air and nitrogen was then determined based on 
these preselected data samples as will be described in the following sections.

\section{Calibration and Geometrical Acceptance}
\label{sect-un-cal}
Before the fluorescence yield can be calculated, the detectors and the DAQ 
system need to be calibrated and the geometrical acceptance of the detectors, 
as they were mounted on the thin target chamber, must be determined.

The fluorescence yield then can be calculated as
\begin{equation}
Y = \frac{1}{N_{e^-}}\cdot \frac{S_{fl}\cdot C}{R_D\cdot G},
\end{equation}
where $S_{fl}$ is the background subtracted and LED corrected fluorescence 
signal in ADC counts, $C$ is the calibration factor converting ADC counts 
into pC, $N_{e^-}$ is the number of electrons in a beam pulse, $G$ is the 
effective geometrical acceptance in m$^{-1}$, and $R_D$ is the 
photon flux responsivity of the detector assembly
in pC~m$^2$/$\gamma$.

As mentioned in the last section only the south signal PMT was used for the 
final result. The calibration constant $C$ for its DAQ channel was measured 
to be 1/3.48 pC/ADC~counts.

\begin{figure}
\includegraphics[angle=-90.,width=\columnwidth]{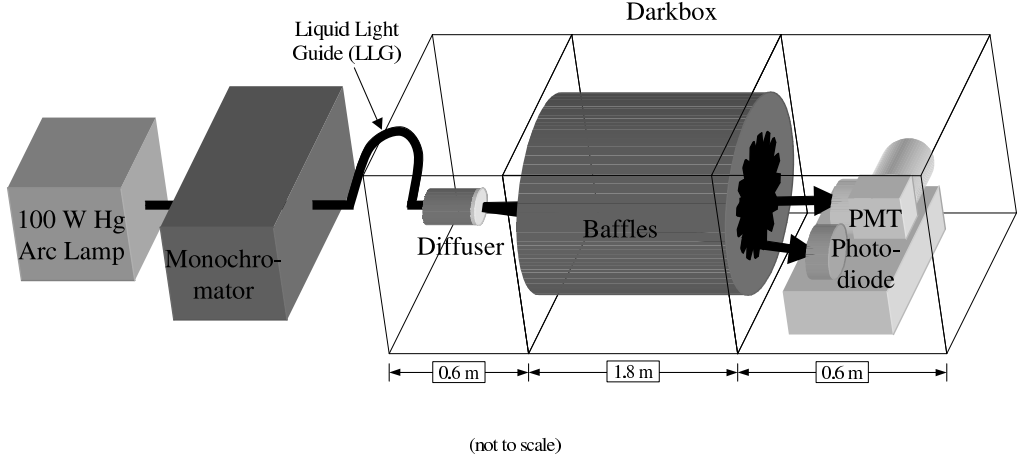} 
\caption{The detector calibration setup.}
   \label{calib-setup}
\end{figure}

The south detector assembly was calibrated after T-461 data taking was 
completed. As can be seen in Figure~\ref{fig-illustration}, the PMT was 
mounted in a brass housing along with a HiRes filter and a mirror. The 
complete unit was calibrated in a standard HiRes calibration setup at the 
University of Utah. A schematic of the calibration setup is shown in 
Figure \ref{calib-setup}. It consisted of a 100~W high-pressure mercury 
arc lamp, a monochromator, 
a light guide and a diffuser as the light source, a 1.8~m light path in a 
baffled foam tube, and a stand with a calibrated silicon 
photo diode installed next to the south PMT unit. The diffuser, the baffled 
light path, the Si photo diode and the PMT unit were enclosed completely in a 
dark box. The south signal PMT detector unit and the silicon photo diode were 
aligned so that they were directly facing the diffuser. During calibration 
the monochromator scanned the wavelength region between 260~nm and 420~nm 
in 1~nm steps. After several monochromator scans were recorded, the PMT and 
photo diode were swapped. In this manner light emitted by the mercury lamp 
was collected by the PMT assembly and the diode.  The current output of each 
detector was measured by a pico ampere meter. 
From the calibration curve of the 
wavelength dependent responsivity of the silicon photo diode in units of 
A/(W/$\mathrm {cm^2}$),
and the measured size of the entrance pupil of the T-461 detector assembly, 
its responsivity could be calculated in units of A/(W/$\mathrm {cm^2}$).
In order to calculate the spectrum weighted photon flux responsivity 
$R_D$ of the detector the wavelength dependent responsivity was folded 
with the normalized fluorescence spectrum of air and nitrogen 
at 760 Torr as reported in Tables 1 and 2 of reference~\cite{Nagano:2004}. 
The fluorescence photon flux responsivity between 300 and 400~nm was found 
to be $R_D=1.47 \times 10^{-6}\mathrm {pC\cdot m^2}/\gamma$ for air
and $R_D=1.66 \times 10^{-6}\mathrm {pC\cdot m^2}/\gamma$ for nitrogen.

The geometrical acceptance $G'$ was calculated as
\begin{equation}
G' =   \frac{W}{4 \cdot \pi \cdot (D-L) \cdot D}
\end{equation}
where $D$ is the distance from the beam axis to the detector unit's aperture,
$L$ is the distance from the beam axis to the slit, and $W$ is the slit width
defining the beam length seen by the detector. $D$, $L$, and $W$ were measured
for the south signal PMT arm and based on the measured numbers $G'$ was 
calculated to be 1/213.8~m$^{-1}$. A small fraction of the energy deposited by the 
beam in the gas was outside the transverse aperture of the optical slit ($\pm$~0.85 cm).
This was evaluated from the Monte Carlo simulations of the process,
leading to a correction of 7\% and an effective geometrical acceptance of
\begin{equation}
G=   G' \cdot l,
\end{equation}
where $l$ is $\sim$ 93\% with a very small pressure dependence.

\section{Resulting Fluorescence Yield}
\label{sect-results}
\begin{figure}
  \includegraphics[width=\columnwidth]{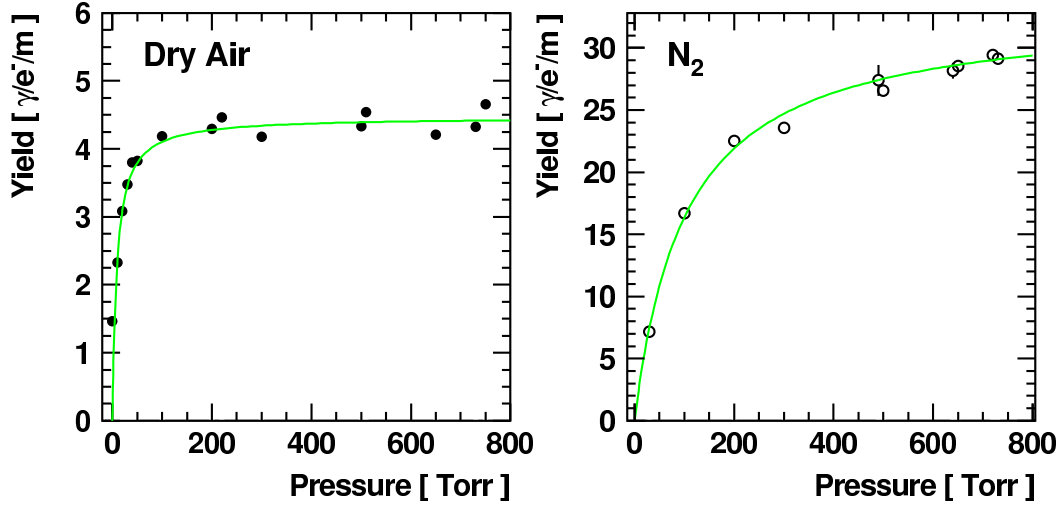}
  \caption{The fluorescence yield curves for air (left plot) and 
N$_2$ (right plot) as measured by the south PMT. The $x$-axis is 
the pressure measured in Torr, and the $y$-axis is the resulting 
fluorescence yield between 300 and 400~nm in photons per electron per
meter ($\frac{\gamma}{\rm e^- m}$).}
  \label{fig-yield fit}
\end{figure}

\begin{figure}
  \includegraphics[width=\columnwidth]{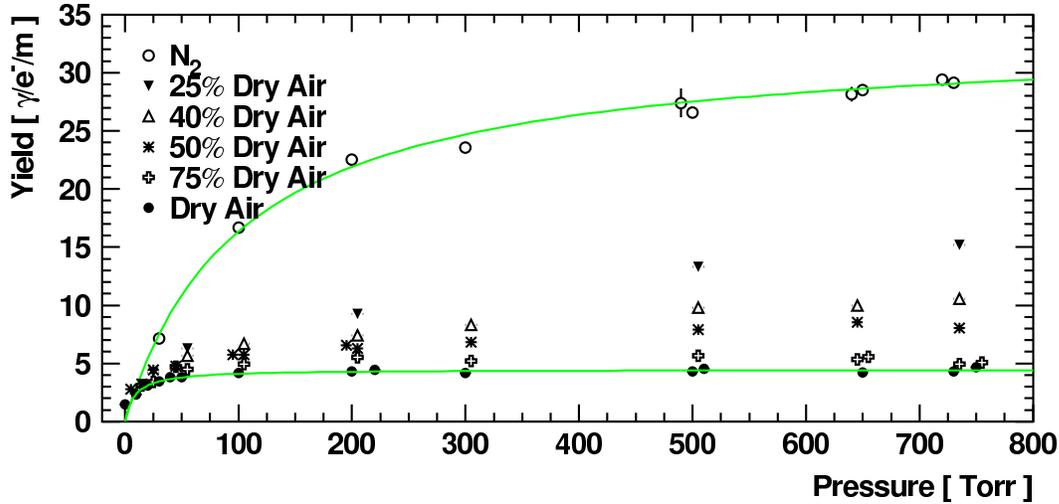}
  \caption{The fluorescence yield curves of air and N$_2$ and various
air-N$_2$ mixtures. The $x$-axis is the pressure measured in Torr, and the 
$y$-axis is the resulting fluorescence yield between 300 and 400~nm in 
$\frac{\gamma}{\rm e^- m}$.}
  \label{fig-yield family}
\end{figure}

Figure \ref{fig-yield fit} shows the calculated yield in units of 
$\frac{\gamma}{\rm e^- m}$ versus the measured pressure in Torr for 
air and nitrogen at 29$^\circ\:$C.
For air, the curve plateaus at about 4.4$\frac{\gamma}{\rm e^- m}$, while
the N$_2$ curve continues to climb with pressure and reaches a yield
approaching seven times that at 800 Torr. 
Figure \ref{fig-yield family} shows the measured
pressure dependent yield of four different air-$\rm N_2$ mixtures framed by
the $\rm N_2$ and air yield curves. Inspired by the fluorescence yield studies
in previous publications~\cite{Bunner:1967,Nagano:2003,Nagano:2004}, where
fits were performed to the measured pressure ($p$) dependent yields of 
individual spectral lines, the function 
\begin{equation}
\label{eq-fy-pdep}
Y(p) = \frac{C}{\frac{1}{p'} + \frac{1}{p}}
\label{eq-y fit}
\end{equation}
was fitted to the air and N$_2$ yield curves using the log likelihood method
with $C$ and $p'$ as freely variable parameters. In the case of an individual
spectral line, $p'$ is the reference pressure specific to the line and $C$ is
the product 
\begin{equation}
C= \frac{dE}{dx}\cdot\frac{1}{RTh\nu}\cdot \Phi^0.
\label{eq-y}
\end{equation}
Here $dE/dx$ is the energy deposited by an electron traveling through the 
gas target, $R$ is the specific gas constant, $T$ is the temperature, $h\nu$ is
the photon's energy, and 
$\Phi^0$ is the fluorescence efficiency of the corresponding
spectral line in the absence of collisional 
quenching~\cite{Bunner:1967,Nagano:2003,Nagano:2004}.
The values for $p'$ derived from the fits are 8.9~$\pm$0.8 Torr for air and 
103$\pm$10 Torr for nitrogen.
The functions, $Y(p)$, 
resulting from the fits were then used to calculate the yields 
in air and nitrogen at 1~atm:
\begin{eqnarray}
  Y(760~{\rm Torr})^{\rm air}= 4.42 \pm 0.73 \frac{\gamma}{\rm e^- m}\nonumber
\end{eqnarray}
and 
\begin{eqnarray}
  Y(760~{\rm Torr})^{\rm N_2}= 29.2 \pm 4.8 \frac{\gamma}{\rm e^- m}
  .\nonumber
\end{eqnarray}

The sources contributing to the overall uncertainty are summarized in table
\ref{table-error}. 
The total uncertainty is 16.6\%.
The statistical uncertainty contributes less than 1\% to 
the total uncertainty. The largest systematic uncertainty is associated with 
the detector calibration (10.5\%). This accounts for the transfer 
uncertainty from the calibration setup in Utah to SLAC, the radiometer
calibration uncertainty of the silicon photo diode, in addition to 
statistical, environmental,
geometrical, and other systematic uncertainties derived from 
a series of calibration measurements. 
The uncertainties of the fluorescence spectrum in air and
nitrogen quoted in Table 1 and 2 of reference~\cite{Nagano:2004} were 
also taken into account.
The second largest contribution of 10\% to the total uncertainty is from the 
beam charge measurement. It was estimated based on systematic studies 
comparing simultaneous measurements of the two different
toroids used during T-461 and FLASH (E-165).
The uncertainty of the effective geometrical acceptance, $G$, of 7\%
is due to a conservative estimate of the uncertainties on the measured slit 
widths and the measured distances to the beam line.
A potential systematic effect from the linearity cut described in 
Section~\ref{sect-selection} was also investigated. The linearity 
requirement was varied between $N_{e^-}<0.5 \cdot 10^9$ and 
$N_{e^-}<1.1 \cdot 10^9$ electrons per bunch resulting in differences of the 
measured yield in air and nitrogen of up to 3\%. The uncertainty of the ADC 
calibration of the south PMT DAQ channel contributes 2\% to the overall
uncertainty. The smallest systematic uncertainty of 1\% is associated with 
the background subtraction. 
\begin{table}
\begin{center}
\begin{tabular}{|l|l|l|l|l|} \hline
\multicolumn{2}{|l|}{Relative error at 760 Torr} &\multicolumn{2}{|c|}{Air and N$_2$} \\ \hline
\multicolumn{2}{|l|}{Statistics}    & \multicolumn{2}{|l|}{$<<$1\%} \\ \hline
\multicolumn{2}{|l|}{Systematics}   & \multicolumn{2}{|l|}{ }      \\ \hline
~~~Detector calibration    & $R_D$     & \multicolumn{2}{|c|}{10.5\%}  \\\hline
~~~Beam Toroid Calibration & $N_{e^-}$ & \multicolumn{2}{|c|}{10\%}   \\ \hline
~~~Geometrical Acceptance  & $G$       & \multicolumn{2}{|c|}{7\%}    \\\hline
~~~Linearity cut           & $S_{fl}/N_{e^-}$ & \multicolumn{2}{|c|}{3\%}\\\hline
~~~ADC calibration         & $C$       & \multicolumn{2}{|c|}{2\%}    \\ \hline
~~~Background subtraction  & $\langle S_{bg}\rangle$ & \multicolumn{2}{|c|}{1\%}    \\ \hline
Total                      & $Y$       & \multicolumn{2}{|c|}{16.6\% }\\ \hline
\end{tabular}
\end{center}
\caption{Summary of statistical and systematic uncertainties in the 
measurement of the total fluorescence yield, $Y$, in air and pure nitrogen at 
760 Torr at 29$^\circ\:$C.}
\label{table-error}
\end{table}

\section{Fluorescence Decay Time}

     The strong saturation of the fluorescence yield in the pressure range up 
to atmospheric (Fig. \ref{fig-yield fit}) is caused by collisional 
de-excitation of the nitrogen molecules. This statistical process
enforces exponential decay times on the excited states. The basic formalism 
is the same as in the previous section and in terms of decay lifetimes, 
$\tau$, may be given as $\tau = \tau_0/(1+p /p')$. Here $p'$ is the same as 
in eq. \ref{eq-fy-pdep} and $\tau_0$ is the decay time in the absence of
collisions, for example at very low pressure.

Most of the wavelengths in this study are transitions in the second positive 
bands, and may be expected to have similar decay properties. The 1N 
transition at 391 nm may be different, but accounts for a small fraction of 
the light at our pressures. In this experiment, decay effects are
averaged over the detected wavelengths. 

\begin{figure}
\begin{center}
\includegraphics*[width=8cm]{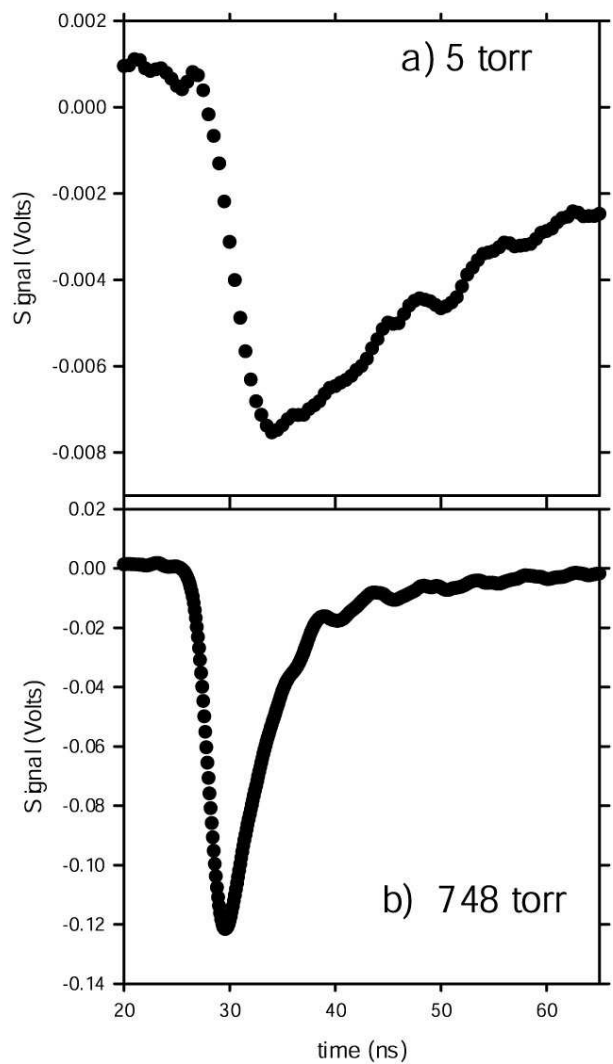}
\end{center}
\caption{Pulse profiles from the photomultiplier for air pressures of 5 and 748 Torr.}
\label{fig-pulse-1}
\end{figure}

An overall dependence of the decay time on pressure was indeed observed. 
As a way of studying it, photomultiplier pulse profiles were recorded at
various gas pressures, using a digital oscilloscope. Typically 16 pulses were 
averaged. The dependence of decay time on pressure was very obvious, as is 
illustrated by the pulses shown in Fig.~\ref{fig-pulse-1} for air at 5 and 748 Torr.

In order to process the data, the small effect of beam induced background in 
the PMT was removed. As described in Section~\ref{sect-selection}, 
this was accomplished by 
using pulse shapes recorded while the test vessel was filled with ethylene, 
corresponding to that 
part of the signal not from the gas fluorescence. Suitably normalized to the 
correct beam intensity, this was subtracted, sample by sample, from the 
fluorescence pulse profiles. The effect was noticeable only below 30 Torr, 
and especially in air which gives smaller signals.

     The next step was to record the PMT intrinsic pulse shape, which was 
measured using an Sr$^{90}$ source to make picosecond long pulses of Cherenkov 
light in the tube face. For computational purposes, this shape was 
parametrized by a well fitting asymmetric probability distribution, 
Pearson IV \cite{Pearson}. (Other functions would have fitted almost 
as well.) The intrinsic shape was then folded with hypothetical light 
pulses which turned on instantly (corresponding to the SLAC picosecond 
electron pulse length) but decayed exponentially. The folding was repeated 
for a range of decay times of the light, and, for each case, the width of 
the folded pulse at half maximum (FWHM) was taken to characterize the pulse 
shape.

\begin{figure}
\begin{center}
\includegraphics*[width=8cm]{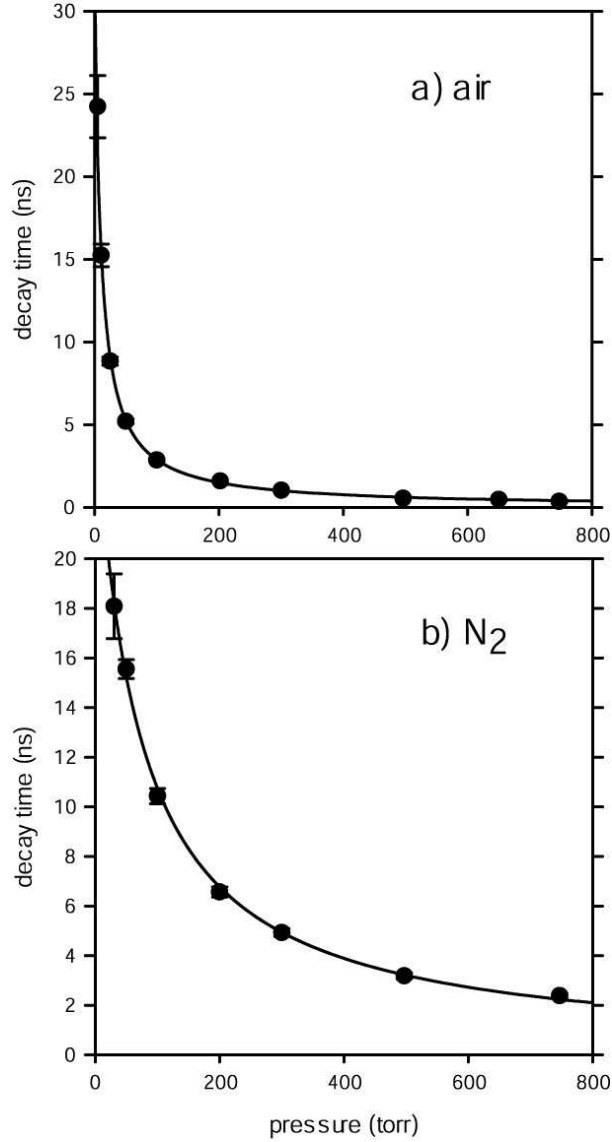}
\end{center}
\caption{Decay time vs. pressure: a) air; b) nitrogen.}
\label{fig-pulse-2}
\end{figure}

     Using this calibration, the measured FWHM values of the actual data 
pulse profiles were transformed to the light decay times. Values for air 
and nitrogen are shown in Fig.~\ref{fig-pulse-2}. The nitrogen data below 30 Torr 
were excluded because of uncertainty about the effect of a small air leak in 
the system. Uncertainties at each point were estimated for the FWHM 
measurements and the background subtraction, and scaled, after fitting, so 
that $\chi2$ was equal to the number of degrees of freedom. Fitting to the 
functional form above allows us to obtain the values for decay times at 
``zero pressure'' (i.e. in the absence of collisional de-excitation) of 
$32.0 \pm 2.2$~ns for air and $26.1 \pm 1.2$~ns for nitrogen. At atmospheric 
pressure, the values found are $0.41 \pm 0.01$~ns for air, and $2.20 \pm 0.06$~ns
for nitrogen. The values for p' are $9.9 \pm 0.7$~Torr for air
and $70 \pm 4$ for nitrogen. The nitrogen value is significantly lower than 
that obtained from the yield curve. This may be related to our assumption 
that all the emission lines can be treated by a single average parameter. 
Differences between them would distort the decay time and yield fits 
differently, especially given the extrapolation below 30 Torr. In future work 
we intend to measure the lines separately.

     In the case of nitrogen, numerous measurements of the ``zero pressure'' 
decay lifetimes have been reported for various wavelength bands, mostly after 
excitation by proton beams. Results, summarized by 
Dotchin et al.~\cite{Dotchin}, show values in the range 34 to 48 ns. A
measurement at $10 ^{-6}$ Torr observed 58 ns~\cite{Blum}. Recently, however, 
Nagano and collaborators~\cite{Nagano:2003}, using electrons to excite the 
nitrogen, report lower values. They give values in six wavelength bands, 
approximately covering the spectral range of our data.
To allow a comparison between the two experiments, we have weighted the decay 
time for each of Nagano's wavebands by its relative intensity and taken a 
weighted average. Their average value for nitrogen is then $27.4 \pm 1.4$ ns. 
A similar average of their results for air is  $30.5 \pm 2$ ns. Our life time
measurements 
agree well with Nagano and collaborators, but not with the earlier proton
beam experiments.

\section{Conclusion and Discussion}
\begin{figure}
  \begin{center}
  \includegraphics*[width=10cm]{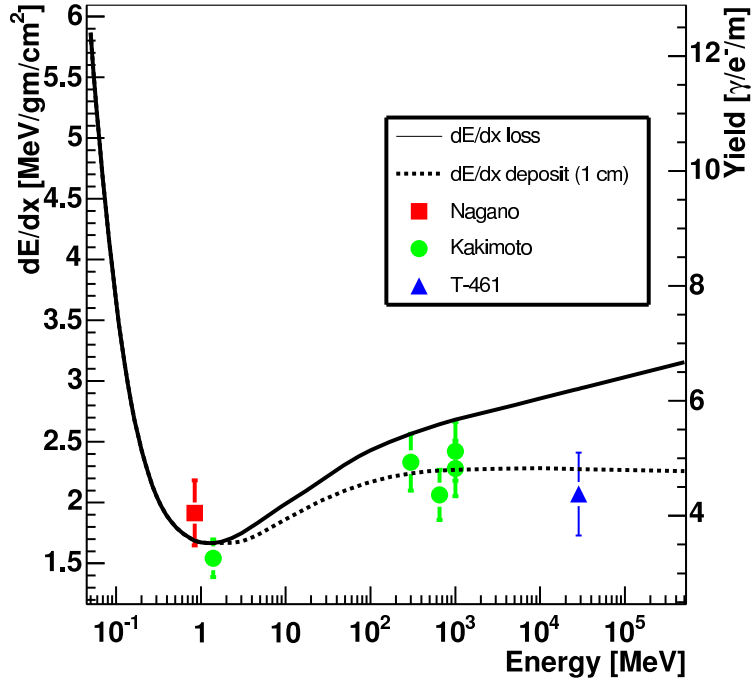}
  \end{center}
  \caption{Comparison of the results from Nagano {\it et al.}\cite{Nagano:2004}, Kakimoto {\it et al.}\cite{Kakimoto:1996}, and T-461.}
  \label{fig-dedx}
\end{figure}

Figure~\ref{fig-dedx} shows the total fluorescence yield between 300 and 400 nm
per electron in air at 760~Torr and 29$^0$C
calculated in this paper 
as well as the yields reported by Kakimoto~\cite{Kakimoto:1996}
and Nagano~\cite{Nagano:2004}. The yields are plotted versus the energy of 
the electrons injected in a thin air target and together with two dE/dx curves.
The line shows the energy loss of electrons in air as calculated based 
on reference~\cite{seltzer:1982}, while the dashed line represents 
the energy deposit of an electron in a 1~cm thick slab of air 
as calculated by GEANT~3~\cite{geant:3}.
The conversion from dE/dx to fluorescence yield was found by performing
a $\chi^2$ fit of the energy deposit to the various measurements. As can be
seen, 
there is good agreement over four decades in electron energy.
The overall uncertainty of the T-461 
result of 16.6\% is 
dominated by systematic effects. Those will be reduced in 
future fluorescence measurements at SLAC by improvements in the calibration of
the light detectors and beam toroid. Among other things, it is planned to
derive an end-to-end calibration of the thin target chamber from
first principles using Rayleigh scattering of a nitrogen laser beam of known
energy sent through the chamber.
The experimental program will also be 
supplemented with a spectrally resolved measurement of the fluorescence light
yield between 290 and 440 nm and an energy dependent yield measurement in
which the electron beam is injected in an air-like target material of 
variable thicknesses to produce confined electromagnetic showers at various
shower depths.

\section{Acknowledgments}
We are indebted to the SLAC accelerator operations staff for their
expertise in meeting the unusual beam requirements, and to personnel of
the Experimental Facilities Department for very professional assistance in
preparation and installation of the equipment.
We also gratefully acknowledge the many contributions from the 
technical staffs of our home institutions.
This work was supported in part by the U.S. Department of Energy under
contract number DE-AC02-76SF00515 as well as by the National Science
Foundation under awards NSF PHY-0245428, NSF PHY-0305516, NSF PHY-0307098,
and NSF PHY-0400053.

%
%
%%%%%%%%%%%%%%%%%%%%%%%%%%%%%%%%%%%%%%%%%%%%%%%%%%%%%%%%%%%%%%%%%%%%%%%%%%%%%%%
% Bibliography
%%%%%%%%%%%%%%%%%%%%%%%%%%%%%%%%%%%%%%%%%%%%%%%%%%%%%%%%%%%%%%%%%%%%%%%%%%%%%%%
%

\end{document}